\documentclass[aip,pop,amsfonts, floatfix, amssymb, amsmath ,reprint ,subscriptadress]{revtex4-1}
\usepackage{times}
\usepackage{bm}
\usepackage{hyperref}
\usepackage{graphicx}
\usepackage{graphics}
\usepackage{epsfig} 
\usepackage{epstopdf}
\usepackage{color}
\usepackage{soul}

\begin{document}

\global\long\def\vp{v_{\parallel}}
\global\long\def\vperp{v_{\perp}}
\global\long\def\rhostar{\rho^{\star}}
\global\long\def\rhotor{\rho_{{\rm {tor}}}}
\global\long\def\rhopol{\rho_{{\rm {pol}}}}

\global\long\def\te{\mathrm{T}_\mathrm{e}}
\global\long\def\ti{\mathrm{T}_\mathrm{i}}
\global\long\def\ne{\mathrm{n}}
\global\long\def\tet{\widetilde{\mathrm{T}}_\mathrm{e}}
\global\long\def\teto{\widetilde{\mathrm{T}}_\mathrm{e,0}}
\global\long\def\tetu{\widetilde{\mathrm{T}}_\mathrm{e,1}}
\global\long\def\tit{\widetilde{\mathrm{T}}_\mathrm{i}}
\global\long\def\net{\widetilde{\mathrm{n}}}

\global\long\def\wc{\mathrm{W}_\mathrm{c}}

\newcommand{\bcomment}[1]{\textcolor{red}{\bf #1}}

\title{Global gyrokinetic simulations of ITG turbulence in the configuration space of the Wendelstein 7-X stellarator}

\author{A.~Ba\~n\'on~Navarro}
\email{alejandro.banon.navarro@ipp.mpg.de}
\affiliation{Max Planck Institute for Plasma Physics, Boltzmannstr.~2, 85748 Garching, Germany}
\author{G.~Merlo}
\affiliation{Oden Institute for Computational Engineering and Sciences, The University of Texas at Austin, Austin, Texas 78712, USA}
\author{G.~G.~Plunk}
\affiliation{Max Planck Institute for Plasma Physics, Wendelsteinstr.~1, 17491 Greifswald, Germany}
\author{P.~Xanthopoulos}
\affiliation{Max Planck Institute for Plasma Physics, Wendelsteinstr.~1, 17491 Greifswald, Germany}
\author{A.~von~Stechow}
\affiliation{Max Planck Institute for Plasma Physics, Wendelsteinstr.~1, 17491 Greifswald, Germany}
\author{A.~Di Siena}
\affiliation{Oden Institute for Computational Engineering and Sciences, The University of Texas at Austin, Austin, Texas 78712, USA}
\author{M.~Maurer}
\affiliation{Max Planck Institute for Plasma Physics, Boltzmannstr.~2, 85748 Garching, Germany}
\author{F.~Hindenlang}
\affiliation{Max Planck Institute for Plasma Physics, Boltzmannstr.~2, 85748 Garching, Germany}
\author{F.~Wilms}
\affiliation{Max Planck Institute for Plasma Physics, Boltzmannstr.~2, 85748 Garching, Germany}
\author{F.~Jenko}
\affiliation{Max Planck Institute for Plasma Physics, Boltzmannstr.~2, 85748 Garching, Germany}
\begin{abstract}

We study the effect of turbulent transport in different magnetic configurations of the Weldenstein 7-X stellarator.  In particular, we performed direct numerical simulations with the global gyrokinetic code GENE-3D, modeling the behavior of Ion Temperature Gradient turbulence in the Standard, High-Mirror, and Low-Mirror configurations of W7-X.  We found that the Low-Mirror configuration produces more transport than both the High-Mirror and the Standard configurations.  By comparison with radially local simulations, we have demonstrated the importance of performing global nonlinear simulations to predict the turbulent fluxes quantitatively.

\end{abstract}
\maketitle
%

%%%%%%%%%%%%%%%%%%%%%%%%%%%%%%%%%%%%%%%%%%%%%%%%%%%%%%%%%%%%%%%%%%%%%%
%%%%%%%%%%%%%%%%%%%%%%%%%%%%%%%%%%%%%%%%%%%%%%%%%%%%%%%%%%%%%%%%%%%%%%
%%%%%%%%%%%%%%%%%%%%%%%%%%%%%%%%%%%%%%%%%%%%%%%%%%%%%%%%%%%%%%%%%%%%%%

\section{Introduction \label{introduction}}

%%%%%%%%%%%%%%%%%%%%%%%%%%%%%%%%%%%%%%%%%%%%%%%%%%%%%%%%%%%%%%%%%%%%%%
%%%%%%%%%%%%%%%%%%%%%%%%%%%%%%%%%%%%%%%%%%%%%%%%%%%%%%%%%%%%%%%%%%%%%%
%%%%%%%%%%%%%%%%%%%%%%%%%%%%%%%%%%%%%%%%%%%%%%%%%%%%%%%%%%%%%%%%%%%%%%

Turbulent transport is one of the major uncertainties in the design of any magnetic fusion device. This is particularly true for stellarators, given the many degrees of freedoms available for the shape of the magnetic geometry~[\onlinecite{Boozer_2015},\onlinecite{Boozer_2019}]. Indeed, with the recent progress in the optimization for neoclassical transport in stellarators, as demonstrated in the planning, construction, and operation of Wendelstein 7-X (W7-X)~[\onlinecite{Wolf_2017},\onlinecite{Pedersen_2018}], it has become clear that turbulent transport is the main mechanism limiting the confinement time~[\onlinecite{Klinger_2019}]. Therefore, in order to further improve stellarators as a viable long-term alternative to tokamaks for future power plants~[\onlinecite{Stroth_1998, Helander_2012, Warmer_2016}], we need to understand, predict, and control turbulent transport.

In this context, the present paper aimed to understand how Ion Temperature Gradient (ITG) turbulence~[\onlinecite{Plunk_2014}], which appears to be one of main micro-instabilities limiting the performance of W7-X in standard scenarios~[\onlinecite{Adrian_2020}], is affected by different magnetic configurations which are experimentally available. This knowledge might provide guidance to prepare and improve future campaigns of W7-X. 

We used the gyrokinetic simulation framework~[\onlinecite{Brizard_2007},\onlinecite{Garbet_2010}] — from which much progress has been made towards understanding turbulence transport in stellarators~[\onlinecite{Helander_2015}] — and performed one of the first global gyrokinetic turbulence studies of W7-X. In particular, we used the gyrokinetic code GENE-3D~[\onlinecite{Maurer_2019}], a newly developed global version of the well established gyrokinetic code GENE~[\onlinecite{Jenko_PoP2000}] for stellarator geometries, to perform linear and nonlinear simulations of W7-X for different magnetic configurations. 

The paper is organized as follows. In Section~\ref{code}, we briefly describe the gyrokinetic code GENE-3D. In Section~\ref{geometry}, the different W7-X magnetic configurations and equilibrium profiles used in this work are introduced.  We then present linear simulation results in Section~\ref{linear}, followed by nonlinear turbulent results in Section~\ref{nonlinear}. A comparison with radially local simulations is performed in Section~\ref{discussion}, demonstrating the importance of global nonlinear simulations to predict turbulent fluxes quantitatively.  Finally, conclusions are given in Section~\ref{conclusions}.

%%%%%%%%%%%%%%%%%%%%%%%%%%%%%%%%%%%%%%%%%%%%%%%%%%%%%%%%%%%%%%%
%%%%%%%%%%%%%%%%%%%%%%%%%%%%%%%%%%%%%%%%%%%%%%%%%%%%%%%%%%%%%%%
%%%%%%%%%%%%%%%%%%%%%%%%%%%%%%%%%%%%%%%%%%%%%%%%%%%%%%%%%%%%%%%

\section{The global gyrokinetic stellarator code GENE-3D\label{code}}

%%%%%%%%%%%%%%%%%%%%%%%%%%%%%%%%%%%%%%%%%%%%%%%%%%%%%%%%%%%%%%%
%%%%%%%%%%%%%%%%%%%%%%%%%%%%%%%%%%%%%%%%%%%%%%%%%%%%%%%%%%%%%%%
%%%%%%%%%%%%%%%%%%%%%%%%%%%%%%%%%%%%%%%%%%%%%%%%%%%%%%%%%%%%%%%

The results presented in this paper are produced with GENE-3D, the global nonlinear version of the GENE code which supports stellarator geometries. GENE-3D has been benchmarked linearly against EUTERPE~[\onlinecite{Kornilov_2004}] in W7-X geometry and nonlinearly against the global version of GENE~[\onlinecite{Goerler_JCP2011}] in tokamak geometry. In addition, GENE-3D has been recently applied to study gyrokinetic turbulence in configurations derived
by novel optimization techniques in W7-X~[\onlinecite{Lobsien_2020}].
In the following section, we outline the main features of the code, but for a detailed description, we refer the reader to Ref.~[\onlinecite{Maurer_2019}].

GENE-3D solves the gyrokinetic Vlasov equation coupled self-consistently to Maxwell's equations on a fixed grid in five-dimensional phase space (plus time), consisting of two velocity coordinates $v_{\parallel}$ (velocity parallel to the magnetic field), and $\mu$ (magnetic moment), and three magnetic field-aligned coordinates $x$, $y$ and $z$, defined as
\begin{align}
x &= \rho_{\rm tor},  \\
y &= \sigma_{B_p} C_y \alpha, \\ 
z &= \sigma_{B_p} \theta^{\star}.
\end{align}
The radial coordinate ($x$) is based on the normalized toroidal flux $\rho_{\rm tor} = \sqrt{\Phi_{\rm tor} / \Phi_{\rm edge}}$, where $\Phi_{\rm tor}$ is the toroidal flux and $\Phi_{\rm edge}$ its value at the last closed flux surface. The bi-normal coordinate ($y$) selects a field line $\alpha= q \theta^{\star} - \varphi$ on a given flux-surface. Here,  $q$ is the safety factor, $\theta^{\star}$ is the poloidal PEST angle~[\onlinecite{LI2016334}] and $\varphi$ is the geometrical toroidal angle.  In the definition of $y$, the constant  $C_y = x_0 / |q_0|$, with $q_0$ being the safety factor at a reference position $x_0$, 
is used to have $y$ as a length rather than an angle-like coordinate.  Finally,  the parallel coordinate ($z$) denotes a position along the magnetic field line. The sign of the poloidal magnetic field $\sigma_{B_p}$ is introduced so that the unit vector in the parallel direction is always in the direction of the magnetic field. 

In these coordinates, the equilibrium magnetic field can be expressed as
\begin{align}
    \mathbf{B_0} = C(x) \nabla x \times \nabla y, 
\end{align}
with $C(x) = (x B_{0}) / ( |q(x)| C_y) $, where $B_{0}$ denotes the magnetic field strength on the magnetic axis.  The magnetic field is a solution of a three-dimensional ideal MHD equilibrium with nested flux surfaces, computed by the Galerkin Variational Equilibrium Code GVEC. In addition to the field and the $q(x)$ profile, GVEC supplies, through an interface, the geometrical information necessary to map the equilibrium to the field-aligned grid. It also provides the mapping into Cartesian coordinates to visualize the simulation data produced by GENE-3D.

To solve the gyrokinetic Vlasov equation numerically, GENE-3D uses the $\delta f$ approach~[\onlinecite{Garbet_2010}]. In this approach, the gyrocenter distribution function $F_{\sigma}$ of species $\sigma$ is split into a background $F_{0 \sigma}$ and a first order perturbation part $F_{1 \sigma}$:
\begin{align}
F_{\sigma}= F_{0 \sigma} + F_{1 \sigma}, \,\, \textrm{with}  \,\, F_{1\sigma}/F_{0 \sigma} <<1, 
\end{align}
where the background distribution function is assumed to be a local Maxwellian in GENE-3D.  

Keeping only first-order terms in the perturbed distribution function, the resulting electrostatic gyrokinetic Vlasov equation solved in GENE-3D reads:
\begin{align}
\begin{split}
     & \frac{\partial{F_{1\sigma}}}{\partial t}  +  v_{\parallel} \hat{\mathbf{b}}_0 \cdot \Gamma_{
  \sigma} - \hat{\mathbf{b}}_0 \cdot  \frac{\mu}{m_{\sigma}} \nabla B_{0} \frac{\partial F_{1\sigma}}{\partial v_{\parallel}}  
 + \left(\mathbf{v}_{\nabla B_0} + \mathbf{v}_c \right) \cdot \Gamma_{\sigma} \\ 
 & +  \mathbf{v}_{E_0}\cdot  \nabla F_{1 \sigma}  +\mathbf{v}_{E_1}  \cdot  \nabla F_{1 \sigma}  
 +  \mathbf{v}_{E_1} \cdot \left[  \nabla F_{0\sigma}  + \mu \nabla B_0  \frac{ F_{0 \sigma}}{T_{0 \sigma}}  \right] \\
  & + ({\bf v}_{\nabla B} + {\bf v}_{c})\cdot\nabla F_{0 \sigma} = C[F_{1\sigma}]\,,
\end{split}
\label{eq: gk}
\end{align}
with
\begin{align}
\Gamma_{\sigma} =  \nabla F_{1\sigma} + \frac{q_{\sigma}}{T_{0 \sigma}} F_{0 \sigma} \nabla \bar{\phi}_1.
\end{align}
The last term on the left-hand side in equation~(\ref{eq: gk}) couples neoclassical and turbulence transport, and it may affect the long-term evolution of the system in the presence of collisions~[\onlinecite{Oberparleiter_2016}]. A linearized Landau-Boltzmann collision operator is currently implemented for $C[F_{1\sigma}]$. In the present paper, neoclassical contributions and collisions are neglected for simplicity.

The drift velocities in equation~(\ref{eq: gk}) are defined as:
\begin{align}
\mathbf{v}_{E_0}   &= \frac{c}{B_0^2} \mathbf{B}_0 \times\nabla\phi_0\,, \label{eq:v_E0} \\
\mathbf{v}_{E_1}   &= \frac{c}{B_0^2} \mathbf{B}_0 \times\nabla\bar{\phi}_1\,, \\
\mathbf{v}_{\nabla B_0} &= \frac{\mu c}{q_\sigma B_0^2} \mathbf{B}_0 \times \nabla B_0\,, \\
\mathbf{v}_c &= \frac{v_{\parallel}^2}{\Omega_\sigma} \left(\nabla \times \mathbf{b}_0 \right)_\perp.
\end{align}
Here,  $\Omega_{\sigma}= (q_{\sigma} B_0) (m_{\sigma} c)$ is the gyrofrequency of a species $\sigma$ with charge $q_{\sigma}$ and mass $m_{\sigma}$, $\hat{\mathbf{b}}_0 = \mathbf{B}_0 / B_0$ and $c$ is the speed of light.
Furthermore,  $\phi_0$ is the equilibrium electrostatic potential which can be employed to consider externally imposed (long-wavelength) radial electric field effects~[\onlinecite{Helander_2008}, \onlinecite{Mishchenko_2012}] and $\bar{\phi}_{1}$ is the gyroaveraged perturbed electrostatic potential, defined as
\begin{align} \label{eq: phibar}
    \bar{\phi}_1(\mathbf{X}) = \frac{1}{2 \pi} \oint  \phi_1(\mathbf{X} + {\bf{r}}(\alpha))\, d\alpha,
\end{align}
where $\mathbf{X}$ is the gyrocenter position and the gyroradius vector ${\bf r}(\alpha)$ is orthogonal to the local magnetic field.

The perturbed electrostatic potential is calculated self-consistently from Poisson's equation, which, when written in terms of the perturbed particle density $n_{1 \sigma}$, reads:
\begin{align} \label{eq: poisson}
\nabla^2_{\perp} \phi_1 = - 4 \pi \sum_{\sigma} q_{\sigma} n_{1\sigma} (\mathbf{x}),
\end{align}
where the left-hand side is neglected in GENE-3D (quasi-neutral limit).  Finally, 
in this work, we treat the electrons as an adiabatic species. 
In this approximation, the adiabaticity relation on a flux-surface reads:
\begin{align}
    \frac{n_{1 e}}{n_{0 e}} = \frac{e}{T_{0 e}} (\phi_1 - \langle \phi_1 \rangle_{\rm{FS}}),
\end{align}
with $e$ being the (positive) elementary charge and $\langle \cdot \rangle_{\rm{FS}}$ denoting a flux surface average~[\onlinecite{Haeseleer_1991}]
\begin{align}
    \langle \;\cdot\; \rangle_{\rm{FS}} = \frac{\partial}{\partial V} \int\limits_V \;\cdot\; dV',
\end{align}
and $V$ being the volume enclosed by that flux surface.

Equation~(\ref{eq: gk}) is solved numerically by discretizing the distribution function on the aforementioned five-dimensional grid. This allows one to reduce the original hyperbolic integro-differential system of equations to a system of ordinary differential equations, which are then explicitly integrated in time. GENE-3D currently uses a 4th order explicit Runge-Kutta (RK4) integrator; the timestep is computed at the initialization and maximized during a run to ensure optimal stability~[\onlinecite{Doerk_2014}]. Because of spatial dependencies in all three spatial coordinates, fourth-order centered finite difference schemes are used to calculate derivatives. This is in contrast to other GENE versions that can rely on spectral (Fourier) methods in certain directions. A zero Dirichlet boundary condition is used in the the radial direction while periodic boundary conditions are applied in the bi-normal direction. The conventional twist-and-shift~[\onlinecite{twist}] boundary condition is applied in the parallel direction after having followed a field line for one poloidal turn. The velocity space is discretized using a regular equidistant grid for $v_{\parallel}$ (assuming as well a zero Dirichlet boundary condition and employing fourth-order centered finite differences)  while a Gauss quadrature scheme, with Gauss-Legendre weights and knots, is used in the $\mu$ direction. 

The Arakawa scheme~[\onlinecite{Arakawa_1966}] is used to evaluate the nonlinear term (the sixth term on the left-hand side of equation~(\ref{eq: gk})) in order to ensure free energy conservation~[\onlinecite{banon_2011}]. 
Finally, the gyroaverage operator, which does not vary during a simulation since it depends only on the background temperature profile and magnetic geometry,  is discretized using finite element in the direction perpendicular to the magnetic field 
using bicubic piecewise polynomials as basis. This allows us to write Poisson's equation as a system of linear equations that is solved using the PETSCs library~[\onlinecite{petsc-web-page,petsc-user-ref, petsc-efficient}]. 
%%%%%%%%%%%%%%%%%%%%%%%%%%%%%%%%%%%%%%%%%%%%%%%%%%%%%%%%%%%%%%%%%%
%%%%%%%%%%%%%%%%%%%%%%%%%%%%%%%%%%%%%%%%%%%%%%%%%%%%%%%%%%%%%%%%%%
%%%%%%%%%%%%%%%%%%%%%%%%%%%%%%%%%%%%%%%%%%%%%%%%%%%%%%%%%%%%%%%%%%

\section{W7-X Magnetic configuration space and equilibrium profiles\label{geometry}}

%%%%%%%%%%%%%%%%%%%%%%%%%%%%%%%%%%%%%%%%%%%%%%%%%%%%%%%%%%%%%%%%%%
%%%%%%%%%%%%%%%%%%%%%%%%%%%%%%%%%%%%%%%%%%%%%%%%%%%%%%%%%%%%%%%%%%
%%%%%%%%%%%%%%%%%%%%%%%%%%%%%%%%%%%%%%%%%%%%%%%%%%%%%%%%%%%%%%%%%

Wendelstein 7-X is the first large, superconducting machine of the HELIAS (HELIcal axis Advanced Stellarator)~[\onlinecite{Nuhrenberg_1986}] type, optimized for strongly reduced neoclassical transport and low bootstrap current. W7-X has a major radius of $R \simeq 5.5$ m, an aspect ratio of $A = R/a \simeq 10$ and $5$ field periods. It has a very flexible coil current system composed of $5$ modular and $2$ planar (tilted) superconducting magnetic coils in each half period. This coil system allows a large variety of magnetic configurations~[\onlinecite{Geiger_2014}, \onlinecite{Dinklage_2018}].

\begin{figure}[b]
\begin{tabular} {c}
\includegraphics[width=0.95\linewidth,angle=0]{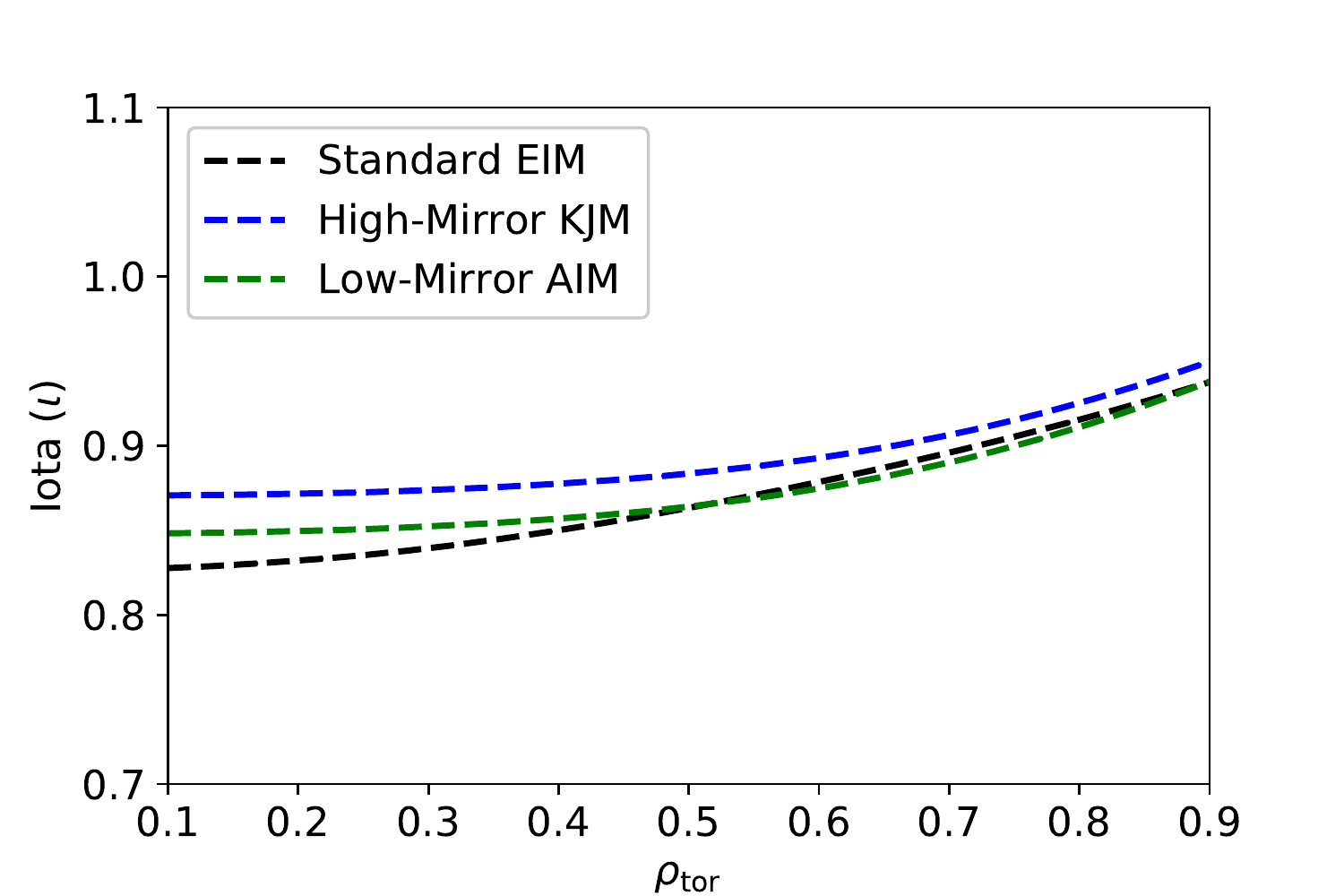} 
\end{tabular}
\caption{\label{fig:iota_profile} Iota profiles for the different W7-X magnetic configurations studied in this paper: Standard (EIM), High-Mirror (KJM) and Low-Mirror (AIM). The High-Mirror configuration has a magnetic mirror ratio of $6 \%$, while is reduced to $3.7 \%$ for the lower-mirror configuration.}
\end{figure}

In this work, we investigate the influence of different magnetic field geometries on Ion Temperature Gradient (ITG) turbulence. For this purpose, we selected three different magnetic configurations: the  Standard configuration (EIM) and two configurations with different magnetic mirror ratios with respect to the Standard configuration; the High-Mirror (KJM), and the Low-Mirror (AIM) configurations. The magnetic equilibria were calculated with the VMEC code~[\onlinecite{vmec_83},\onlinecite{vmec_91}].  The change in the magnetic mirror ratio slightly changes the rotational transform (see figure~\ref{fig:iota_profile}). 

%%%%%%%%
%%%%%%%

\begin{figure}[t]
\begin{tabular} {c}
\includegraphics[width=0.95\linewidth,angle=0]{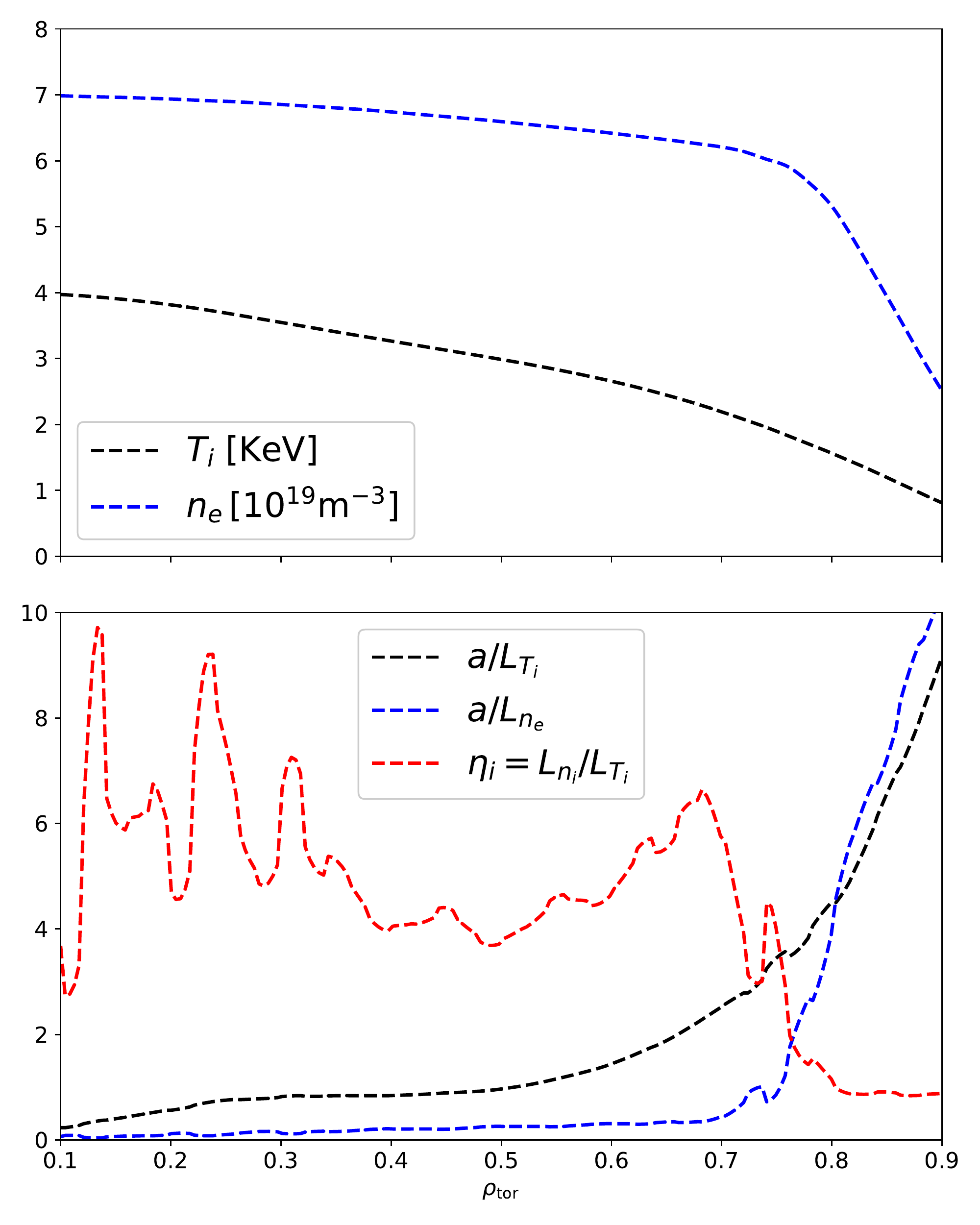} 
\end{tabular}
\caption{\label{fig:profiles} (Top)  Ion temperature and electron density profiles  and (bottom) their respective logarithmic gradients and $\eta_i$ profiles used in this work. }
\end{figure}

To isolate the effect of the magnetic configuration on ITG turbulence, we used the same equilibrium profiles in all previously described magnetic geometries. We also assumed equal ion and electron temperatures. The profiles and their respective normalized logarithmic gradients, with the latter defined as $a / L_{X} = - a \, d \ln X$ where $X$ is the ion temperature ($T_i$) or the electron density ($n_e$), are shown in figure~\ref{fig:profiles}. For simplicity, the profiles were not taken from an actual W7-X discharge, but were produced with the DKES code~[\onlinecite{Hirshman_1986},\onlinecite{rij_1989}] assuming
an idealized W7-X scenario characterized by a broad range of radial positions in which
the ion logarithmic temperature gradient is large and the ratio $\eta_i = L_{n_i} / L_{T_i}$ is greater than one, as shown in figure~\ref{fig:profiles}. In such a scenario, we expect ITG modes to be strongly driven~[\onlinecite{Plunk_2014}], and trapped electron physics,  which is neglected in the adiabatic electron response employed in the simulations, to be less important in capturing the dependence of ITG turbulence on the different magnetic configurations correctly~[\onlinecite{Alcuson_2020}].

%%%%%%%
%%%%%%%

%%%%%%%%%%%%%%%%%%%%%%%%%%%%%%%%%%%%%%%%%%%%%%%%%%%%%%%%%%%%%%%%%%%%%%
%%%%%%%%%%%%%%%%%%%%%%%%%%%%%%%%%%%%%%%%%%%%%%%%%%%%%%%%%%%%%%%%%%%%%%
%%%%%%%%%%%%%%%%%%%%%%%%%%%%%%%%%%%%%%%%%%%%%%%%%%%%%%%%%%%%%%%%%%%%%%

\section{Linear Global ITG Simulations \label{linear}}

%%%%%%%%%%%%%%%%%%%%%%%%%%%%%%%%%%%%%%%%%%%%%%%%%%%%%%%%%%%%%%%%%%%%%%
%%%%%%%%%%%%%%%%%%%%%%%%%%%%%%%%%%%%%%%%%%%%%%%%%%%%%%%%%%%%%%%%%%%%%%
%%%%%%%%%%%%%%%%%%%%%%%%%%%%%%%%%%%%%%%%%%%%%%%%%%%%%%%%%%%%%%%%%%%%%%

%%%%%%%
\begin{figure}[t]
\begin{tabular} {c}
\includegraphics[width=0.95\linewidth,angle=0]{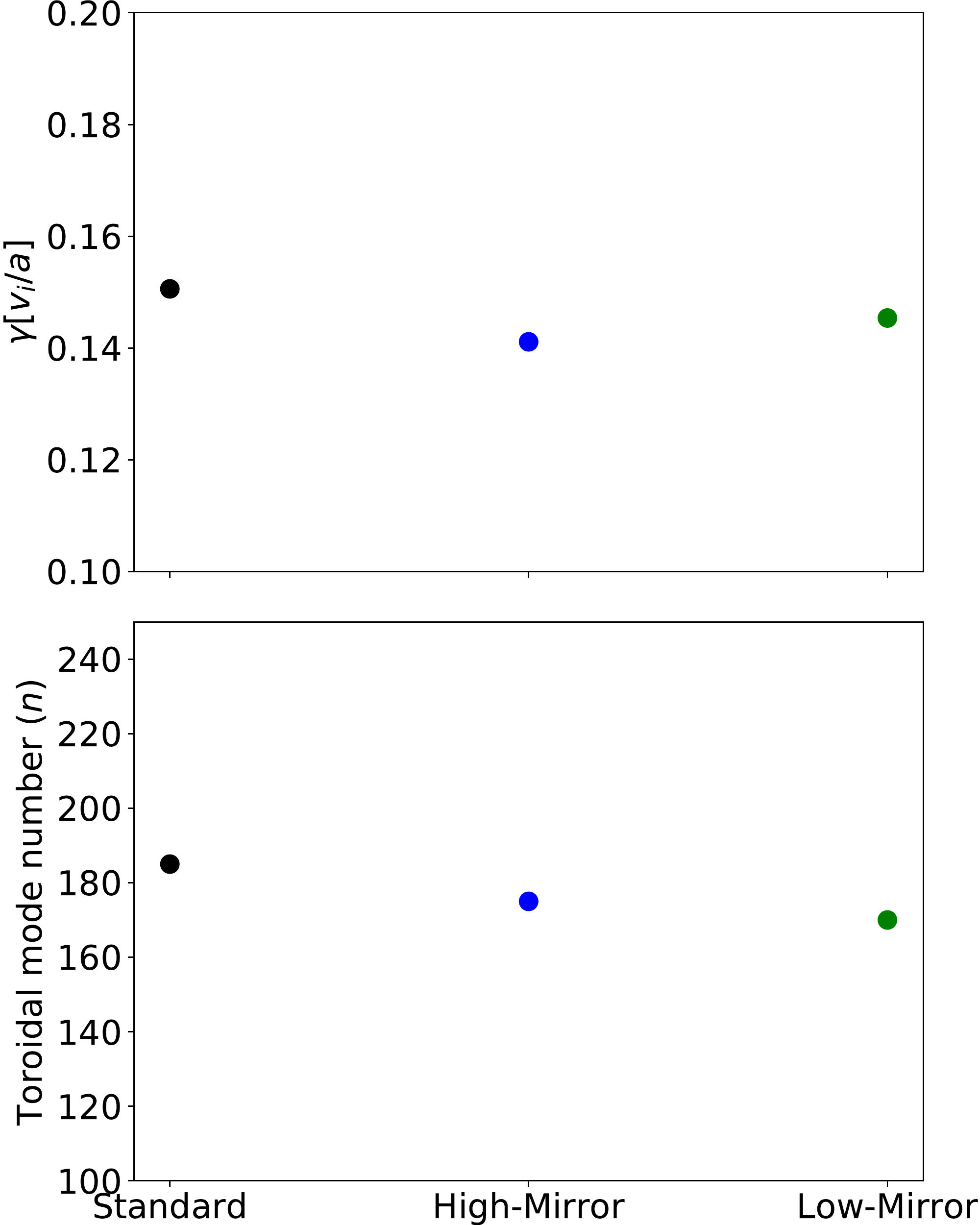}  
\end{tabular}
\caption{\label{fig:growth_rate} (Top) Growth rate ($\gamma$) and  (bottom) peak toroidal mode number ($n$) for the different magnetic configurations.}
\end{figure}
%%%%%%%

Before performing turbulence simulations, we first want to characterize the linear modes present in the different W7-X magnetic configurations. Therefore, in this section, we compare the growth rate, the toroidal mode numbers, and the mode structure for the different cases considered in this paper. The numerical setup for linear simulations includes the following: neglect of the nonlinear term so that only ITG modes are captured; a radial domain from $\rho_{\rm tor} \in [0.1, 0.9]$ with zero Dirichlet boundary condition;  a buffer zone of $5 \%$ of the radial domain with a Krook damping operator with a coefficient of $1.0~v_i/a$ — significantly larger than the expected growth rate of the instability — at each side of the radial domain to avoid numerical instabilities close to the boundaries; and finally, a grid resolution of $\{ 192, 256, 128, 48, 12 \}$ in the $\{ x,y,z,v_{\parallel}, \mu\}$ directions, respectively. Here, $v_i = \sqrt{T_{i,0}/m_i}$, where $T_{i,0}$ is the ion temperature at the reference position $x_0$ (taken here at $x_0 = 0.5$) and $m_i$ is the ion mass.

We display in figure~\ref{fig:growth_rate} the growth rates ($\gamma$) and peak toroidal mode numbers ($n$ is the mode number corresponding to the peak in the Fourier spectrum of $\phi_1$) for the different magnetic configurations. We observe that the growth rates and the peak toroidal mode numbers are very similar for all the different W7-X configurations.
In particular, the ITG mode is characterized by a maximum growth rate between $0.14-0.16~v_i /a$ and a peak toroidal mode number around $n\approx180$.

\begin{figure}[t]
\begin{tabular} {c}
\includegraphics[width=1.0\linewidth,angle=0]{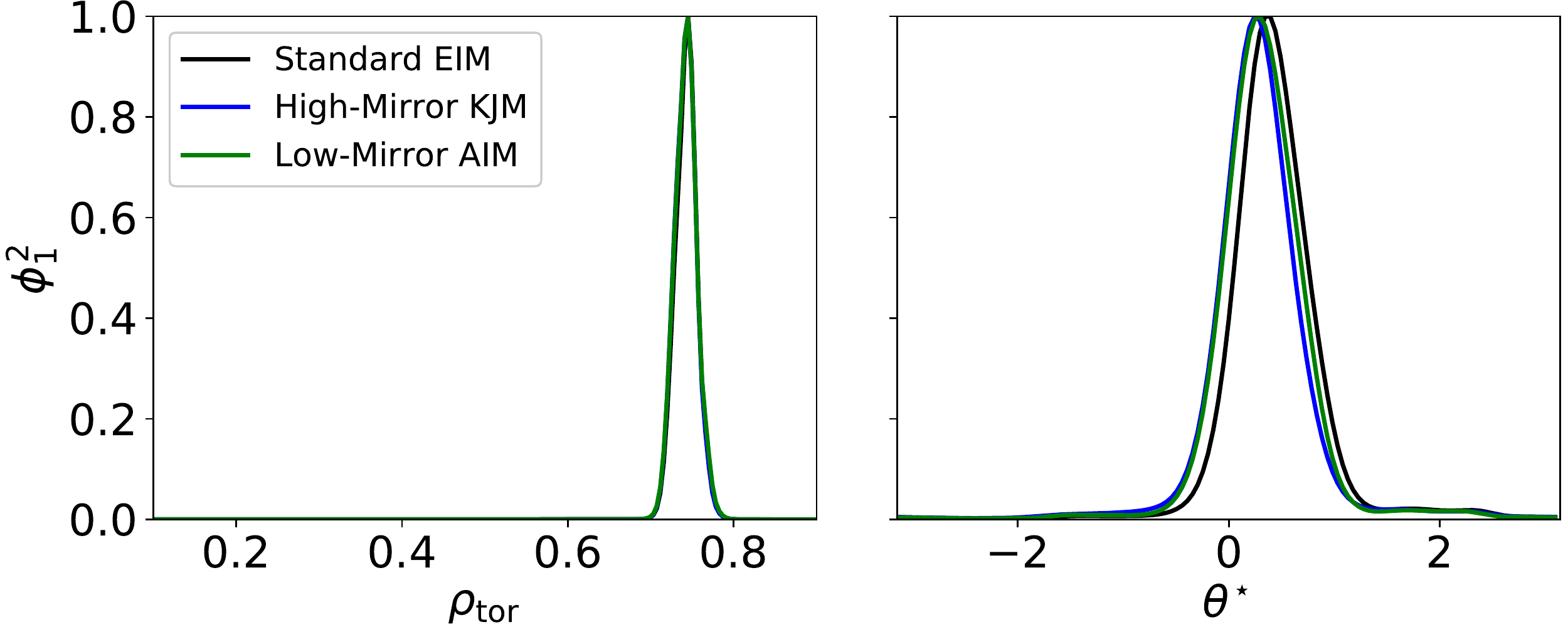} \\ 
\includegraphics[width=0.95\linewidth,angle=0]{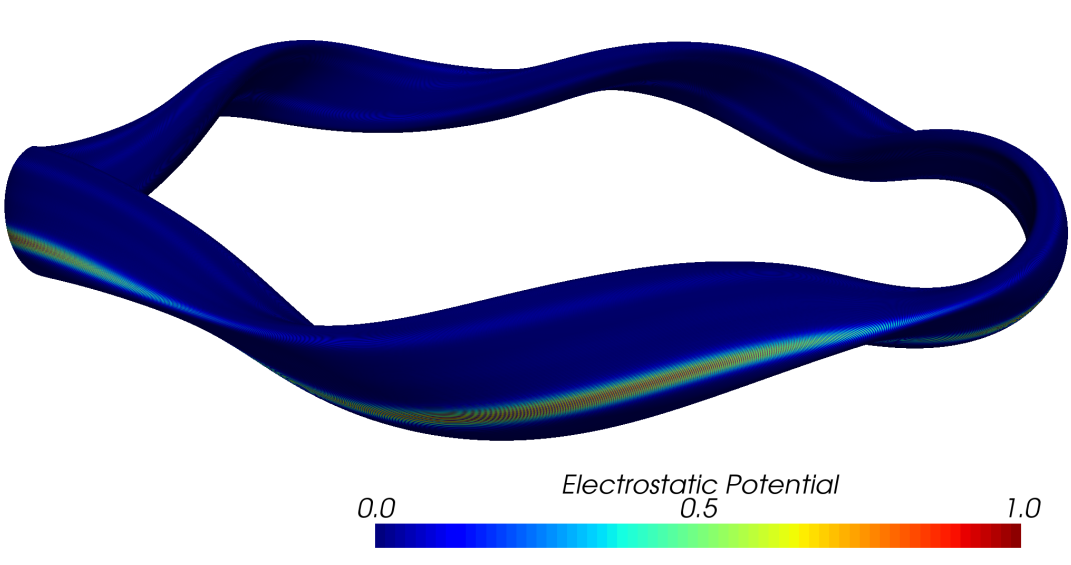} 
\end{tabular}
\caption{\label{fig:slices_phi} (Top left) Square amplitude of the electrostatic potential $\phi_{1}^2$ vs.
the radial coordinate ($\rho_{\rm tor}$)  and (top right) vs. the PEST poloidal coordinate ($\theta^{\star}$)
averaged over the remaining coordinates. (Bottom) Three dimensional representation of the square amplitude of the normalized electrostatic potential at $\rho_{\rm tor}=0.75$ for the Standard configuration.}
\end{figure}

We also found that the mode structure is very similar for all configurations. We show this in figure~\ref{fig:slices_phi} by comparing the square amplitude of the electrostatic potential $\phi_{1}^2$ vs. the radial coordinate (top left)  and vs. the PEST poloidal angle (top right) -- averaged over the remaining coordinates. The mode is localized around the outboard mid-plane ($\theta^{\star} \approx -0.5$) and peaks at $\rho_{\rm tor} \approx 0.75$. This radial location is at a position where $a/L_{T_i}$ is large, and $\eta_{i} = L_{n_i} / L_{T_i}$ is at a local maximum (see figure~\ref{fig:profiles}), thus, favoring strongly unstable ITG modes. Finally, the three-dimensional mode structure in this flux-surface is also shown in figure~\ref{fig:slices_phi}~(bottom) but only for the Standard EIM configuration to avoid redundancy. We observe that the mode is highly localized in the flux-surface, which agrees with previously published linear results of \mbox{W7-X} by EUTERPE~[\onlinecite{Kornilov_2004}, \onlinecite{Kornilov_2005}, \onlinecite{Riemann_2016}].

To summarize this section, we found that ITG modes — modeled with an adiabatic electron response — do not vary significantly for the different W7-X configurations studied in the present paper. They are characterized by large toroidal mode numbers ($n\approx180$), and their mode structure is highly localized both radially and in the flux-surface. 

%%%%%%%
%%%%%%%

%%%%%%%%%%%%%%%%%%%%%%%%%%%%%%%%%%%%%%%%%%%%%%%%%%%%%%%%%%%%%%%%%%%%%%%%%%%%%%%%
%%%%%%%%%%%%%%%%%%%%%%%%%%%%%%%%%%%%%%%%%%%%%%%%%%%%%%%%%%%%%%%%%%%%%%%%%%%%%%%%
%%%%%%%%%%%%%%%%%%%%%%%%%%%%%%%%%%%%%%%%%%%%%%%%%%%%%%%%%%%%%%%%%%%%%%%%%%%%%%%%

\section{Nonlinear Global ITG Simulations \label{nonlinear}}

%%%%%%%%%%%%%%%%%%%%%%%%%%%%%%%%%%%%%%%%%%%%%%%%%%%%%%%%%%%%%%%%%%%%%%%%%%%%%%%%
%%%%%%%%%%%%%%%%%%%%%%%%%%%%%%%%%%%%%%%%%%%%%%%%%%%%%%%%%%%%%%%%%%%%%%%%%%%%%%%%
%%%%%%%%%%%%%%%%%%%%%%%%%%%%%%%%%%%%%%%%%%%%%%%%%%%%%%%%%%%%%%%%%%%%%%%%%%%%%%%%

Having characterized linear ITG modes for all configurations, we proceed to the study of  nonlinear simulations of ITG turbulence. There are two ways of performing global nonlinear simulations and reaching a quasi-steady state: the flux-driven approach, in which fixed sources (e.g., of energy)  are used and cause the profiles to slowly evolve toward a quasi-steady state; and 
the gradient-driven approach,  which employs sources and sinks to keep the plasma profiles close to the initial ones, so that the turbulence drive, characterized by the pressure gradients,  is maintained constant during the simulation.  The flux-driven setup can be considered closer to the experimental situation, but it requires longer simulations. The gradient-driven approach needs shorter runs to reach a quasi-steady state and is, therefore, less computational demanding.  In this work, we adopted the gradient-driven approach. The simulation setup used is as follows: the resolution used is the same as in the linear case, $\{ x,y,z,v_{\parallel}, \mu\} = \{ 192, 256, 128, 48, 12 \}$ -- we have performed convergence studies to ensure the validity of this choice; a Krook-type heat source with a relaxation rate of $0.02 \, v_i/a$  was used to maintain the temperature profile near its initial value; and finally, the results of the simulations were time-averaged over about $1000 \, a / v_i$ after the turbulence reaches a 
quasi-steady state. Each of the simulations presented in the following section required about $0.4$ million CPU-hours on the SkyLake Marconi Supercomputer.

 %%%%%%%
\begin{figure}[t]
\begin{tabular} {c}
\includegraphics[width=0.95\linewidth,angle=0]{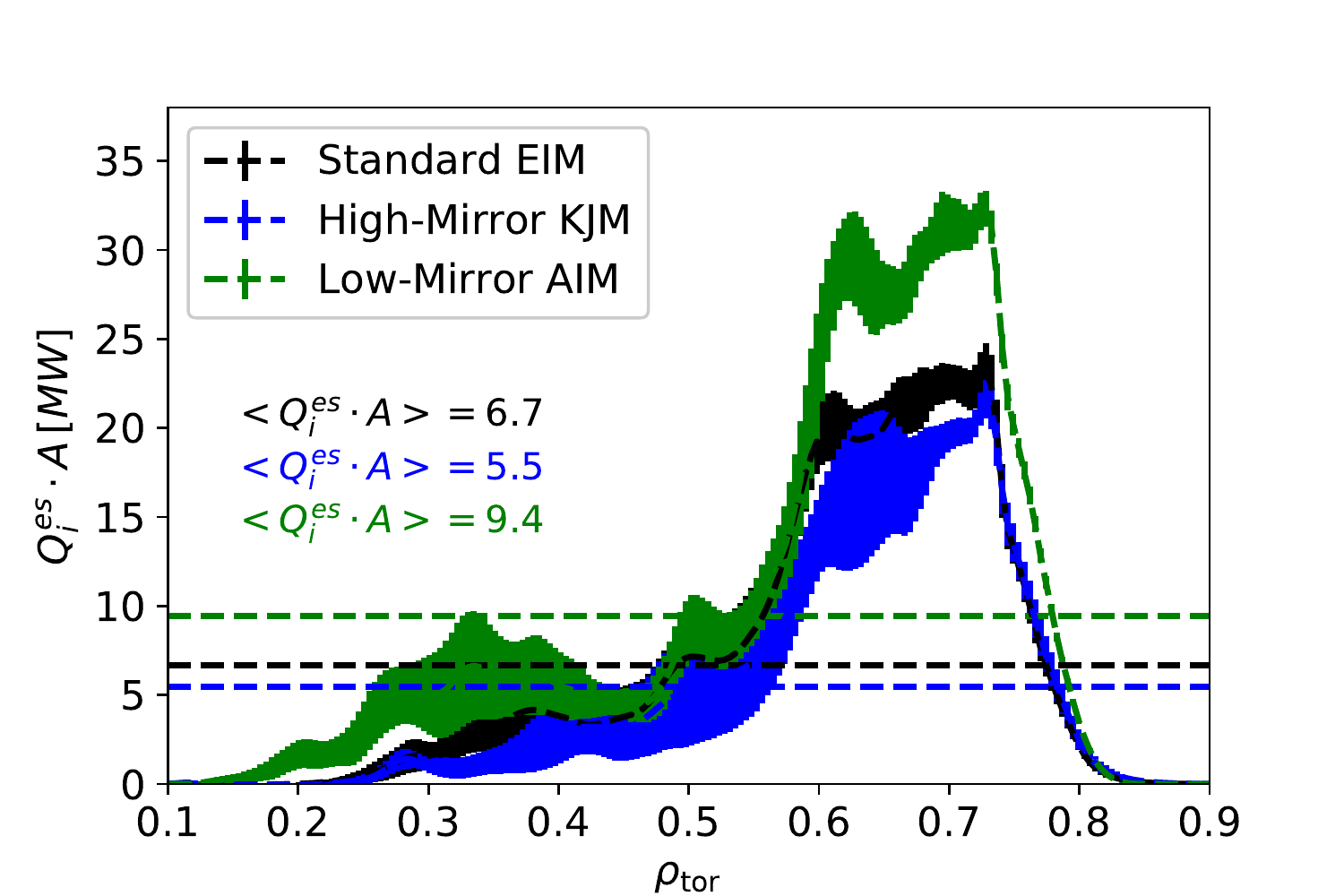} 
\end{tabular}
\caption{\label{fig:heat_profiles} Comparison of the radial profile of the electrostatic ion heat flow for the different magnetic configurations.  The shaded region corresponds to an estimate of the standard deviation of the set of means of consecutive temporal sub-domains of the saturated state and the horizontal dashed line shows its radial average whose value is also indicated.}
\end{figure}
%%%%%%%

Despite similar linear characteristics as the others, the Low-Mirror configuration produces the highest amount of  turbulent transport. This is shown in figure~\ref{fig:heat_profiles} by comparing the radial profile of the time averaged electrostatic ion heat flow ($Q^{\rm es}_i   \cdot A$) for the different magnetic configurations. Here, $Q^{\rm es}_i$ is the flux-surface averaged electrostatic ion heat flux defined as:
 \begin{align}
Q^{\rm es}_{i}  =  \bigg \langle \int \frac{1}{2} m_i v^2 F^{pc}_{1,i}\, \mathbf{v}_{E_1} d \mathbf{v}\, \cdot \nabla x \bigg \rangle_{\rm FS}\,,
\end{align}
where $F^{pc}_{1i}$ is the perturbed part of the ion particle distribution function,
and $A$ is the flux-surface area.  
In this figure, the shaded region corresponds to an estimate of the standard deviation of the set of means of consecutive temporal sub-domains of the saturated state; and the horizontal dashed lines show  their radial average over the whole region.  In particular, we observe that the Low-Mirror configuration produces approximately $1.7$ times more radially averaged heat flow than the High-Mirror configuration and $1.4$ times higher than the Standard configuration.
The difference in the heat fluxes is mainly localized in the region around $\rho_{\rm tor} = 0.65 - 0.75$. In this region, the heat flow can reach values around $30$ MW for the Low-Mirror configuration and around $20$ MW for the other two configurations.  These values are larger than the currently available heating power of W7-X.
Although a quantitative estimate of the flux amplitudes would require a more comprehensive plasma description, e.g. kinetic electron physics, and/or electromagnetic fluctuations, these results may suggest that the profiles used in this work could not  be achieved experimentally in W7-X. 

\begin{figure}[t]
\begin{tabular} {c}
\includegraphics[width=0.95\linewidth,angle=0]{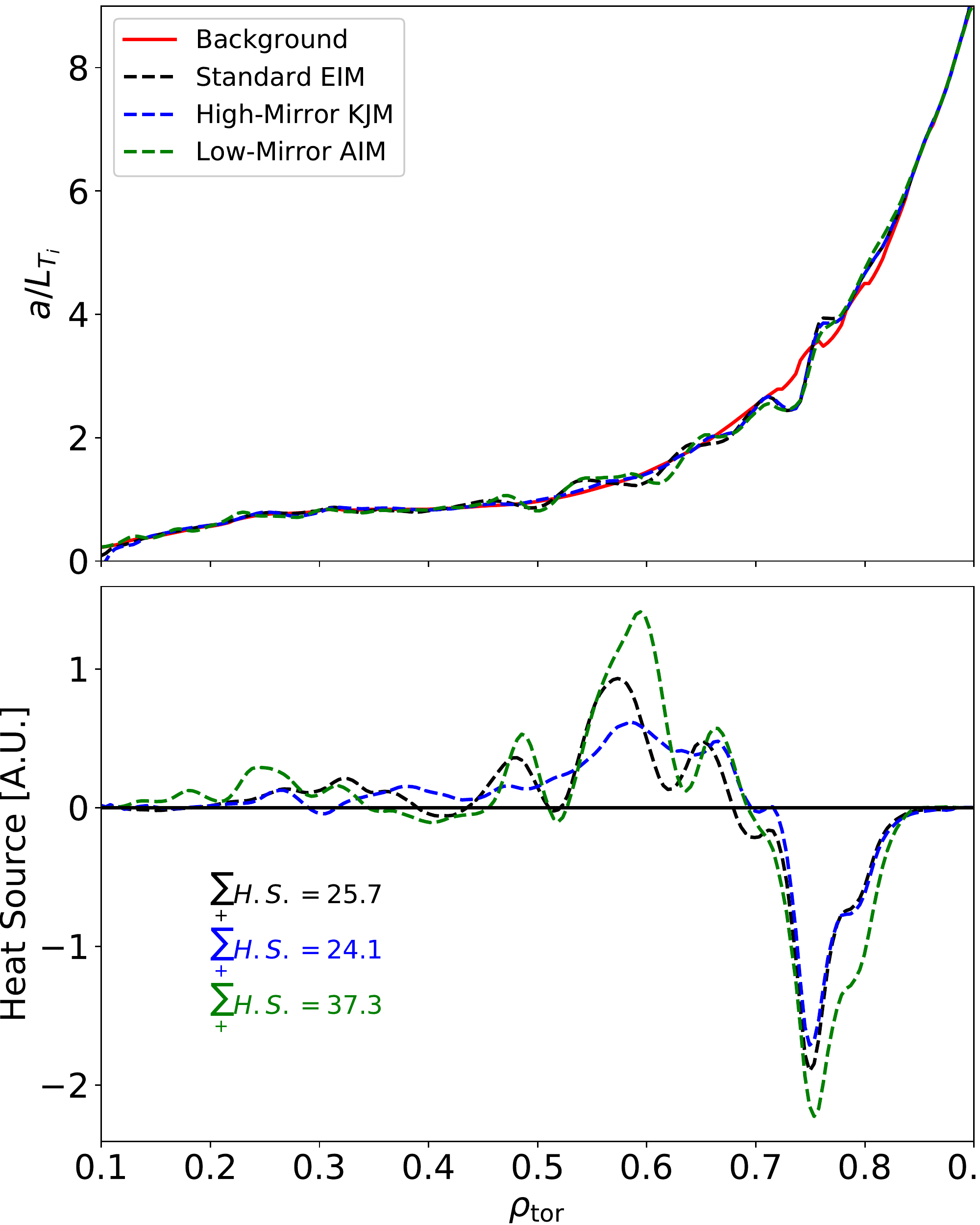}
\end{tabular}
\caption{\label{fig:profiles_turbulence} (Top) Comparison of the logarithmic ion temperature profile and (bottom) heat source profile.
The sum of the positive contributions of the heat source term $\sum_{+} H. S$ is also indicated.}
\end{figure}

Note that for the Low-Mirror configuration to produce more heat flux, while maintaining the profiles, more heat must be injected by the Krook-type heat source term. This is indeed the case as shown in figure~\ref{fig:profiles_turbulence}, where we plot the profiles of the logarithmic ion temperature gradient (top) and the heat source term (bottom). We observe that the ion gradients are well maintained during the simulation for all cases. However, the net heat source input, given by the sum of the positive contributions of the heat source term, is indeed higher for the Low-Mirror configuration.

\begin{figure}[t]
\begin{tabular} {c}
\includegraphics[width=0.95\linewidth,angle=0]{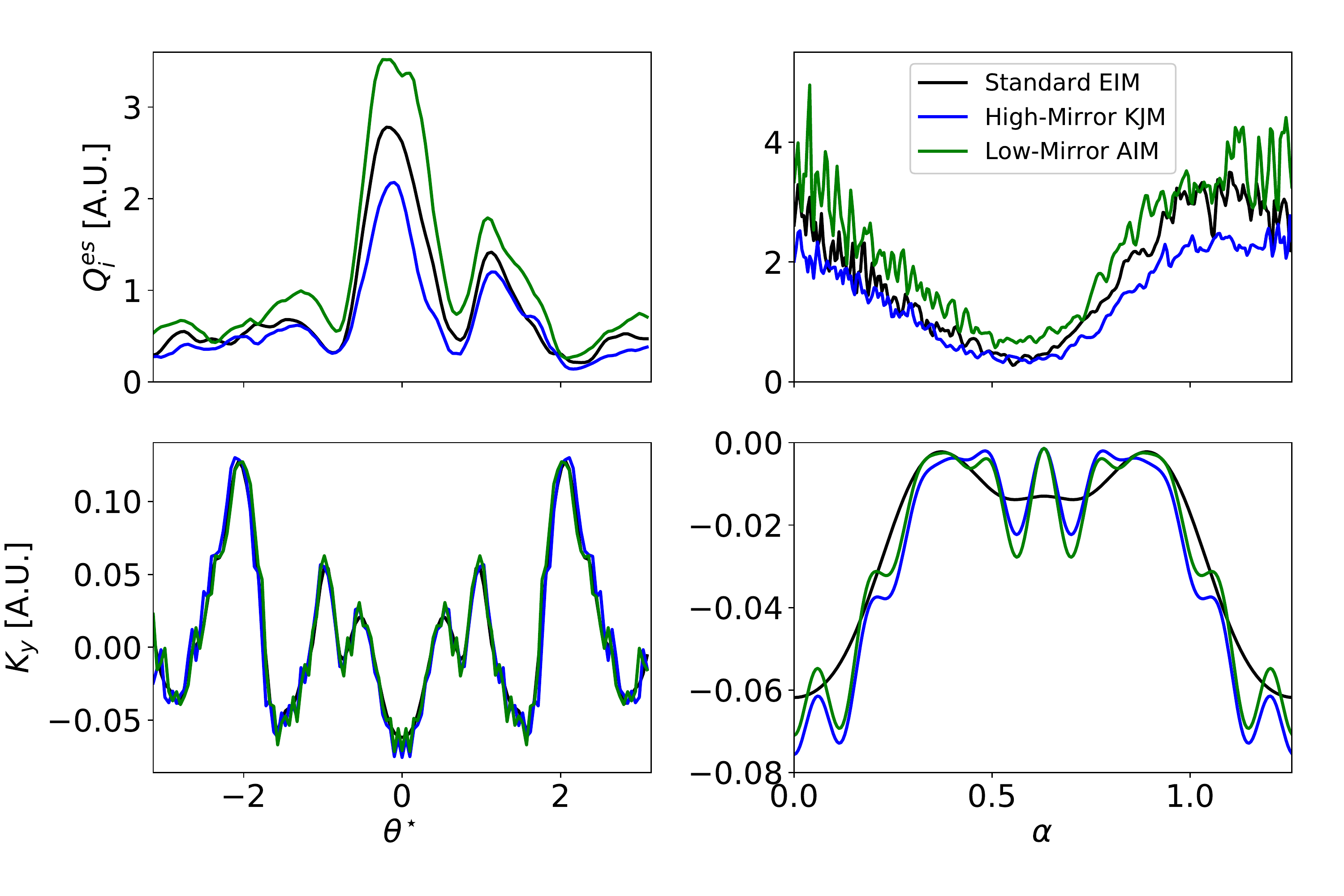}
\end{tabular}
\caption{\label{fig:heat_k_slices} Time averaged electrostatic ion heat fluxes vs. $\theta^{\star}$ (for $\alpha = 0$) (top left)  and  vs. the
field-aligned angle $\alpha$ (for $\theta^{\star} = 0$) (top right)  at $\rho_{\rm tor} = 0.7$. Bi-normal curvature $K_y$ vs. $\theta^{\star}$  (bottom left)  and  vs. $\alpha$  (bottom right) at the same radial position.}
\end{figure}

Although the amplitudes are different between the configurations, the spatial structure of the heat flux is very similar between them. As we show in figure~\ref{fig:heat_k_slices}, the spatial structure of the heat flux at $\rho_{\rm tor} = 0.7$ is maximal around $\theta^{\star} \approx 0$ and at $\alpha \approx 0$ (and $2 \pi/5$ due to periodicity) for all cases. We then compare this structure with that of the bi-normal curvature,  defined as $K_y = - \left( \mathbf{B_0} \times \mathbf{\nabla} B_0\right) \cdot \hat{y} / {B_0^2}$. As expected by the fact that ITG modes are driven unstable by negative values of bi-normal curvature, we observe that the heat flux is at a maximum in the regions of bad curvature, i.e.,  where $K_y$ is at a minimum. 
Taking into account that at this position the bi-normal curvature has a very similar structure (see figure~\ref{fig:heat_k_slices}(bottom)) for all the magnetic configurations, the differences in transport must come from a different geometrical term.

\begin{figure}[t]
\begin{tabular} {c}
\includegraphics[width=0.95\linewidth,angle=0]{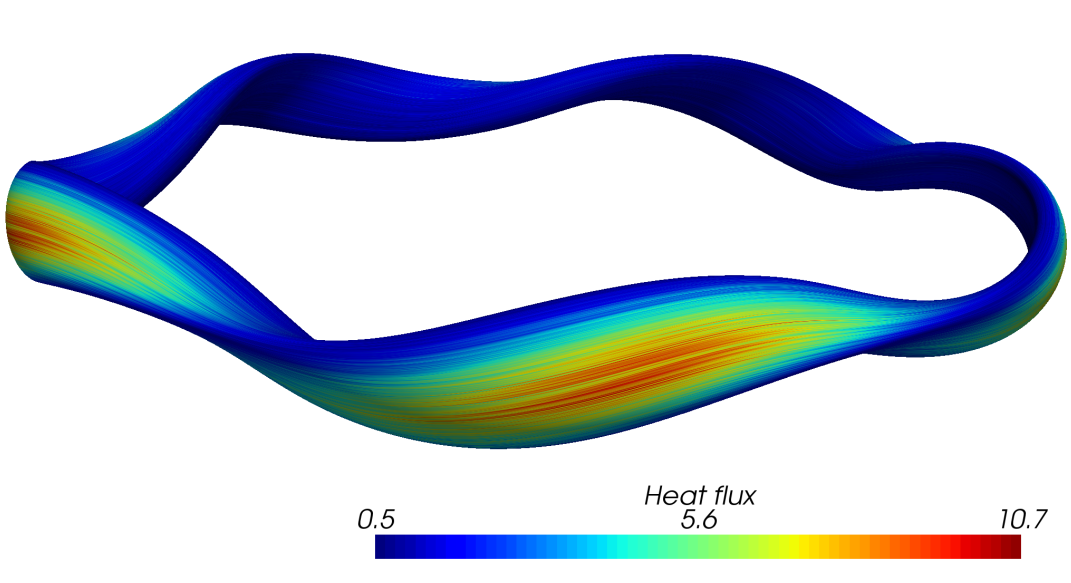} 
\end{tabular}
\caption{\label{fig:heat_3d} Three-dimensional representation  of the time averaged electrostatic  ion heat flux at   $\rho_{\rm tor} = 0.7$ for the Standard configuration.}
\end{figure}

\begin{figure}[t]
\begin{tabular} {c}
\end{tabular}
\includegraphics[width=0.95\linewidth,angle=0]{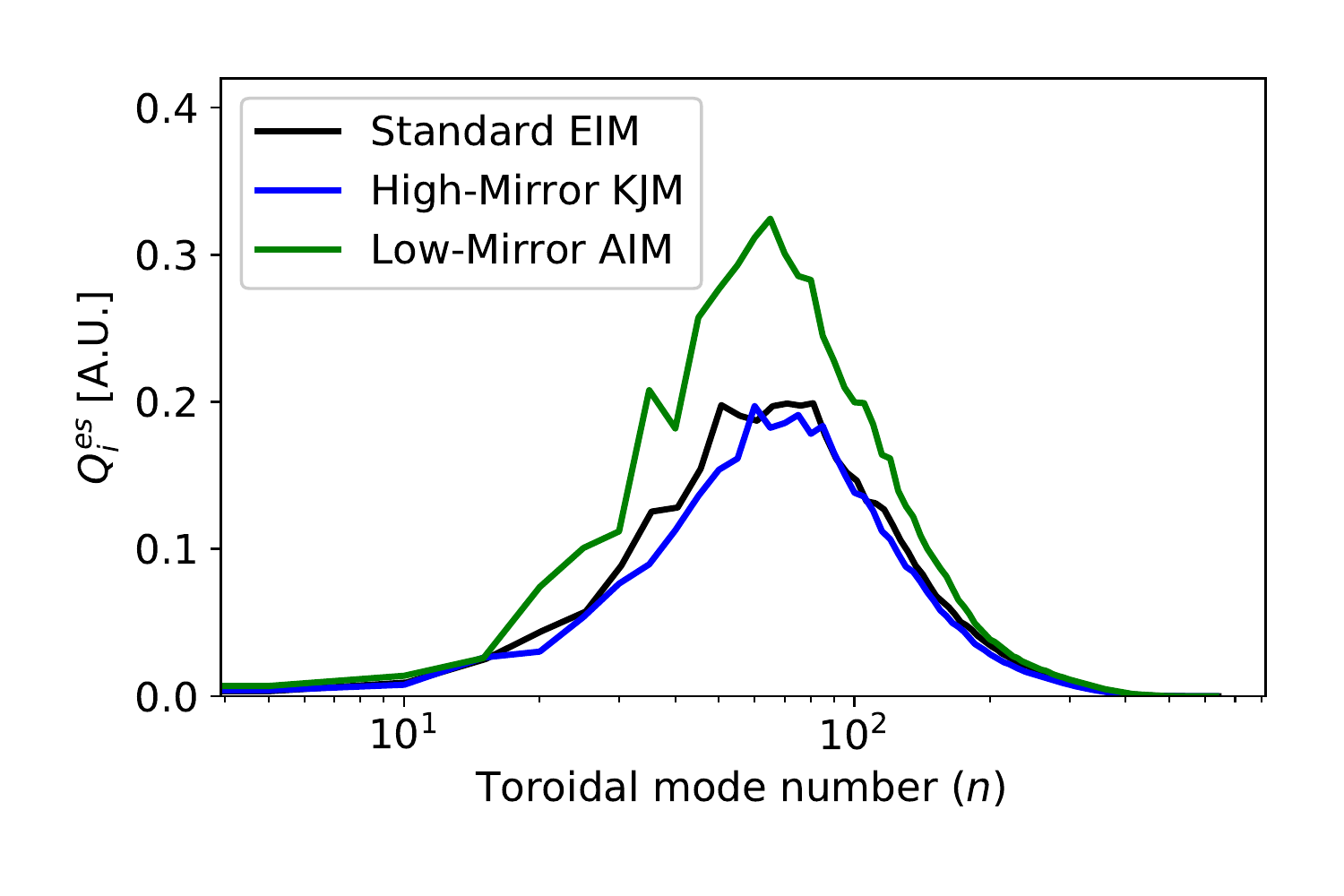} \\
\caption{\label{fig:heat_flux_spectra} Comparison of the ion heat flux spectra at $\rho_{\rm tor} = 0.7$ for the different magnetic configurations.}
\end{figure}

Furthermore, in order to compare with the linear case, we display in figure~\ref{fig:heat_3d} the three-dimensional representation of the time average heat flux, but only for the Standard configuration. We observe that most transport is localized in a narrow region at the outboard mid-plane. This localization is, however, broader than the one presented by the linear mode structure (see figure~\ref{fig:slices_phi}). This difference can be explained by analyzing the ion heat flux spectra at $\rho_{\rm tor} = 0.7$. As shown in figure~\ref{fig:heat_flux_spectra}, the heat flux spectra are dominated by a broad range of toroidal modes with a maximum around $n\approx70$. This maximum is therefore a larger scale mode than the most dominant mode in linear simulations ($n\approx180$).  In fact, the most unstable mode in linear simulations does not contribute significantly to transport. This indicates that it is crucial to perform nonlinear simulations to characterize the system correctly.

%

%%%%%%%
%%%%%%%

\begin{figure}[t]
\begin{tabular} {c}
\includegraphics[width=0.95\linewidth,angle=0]{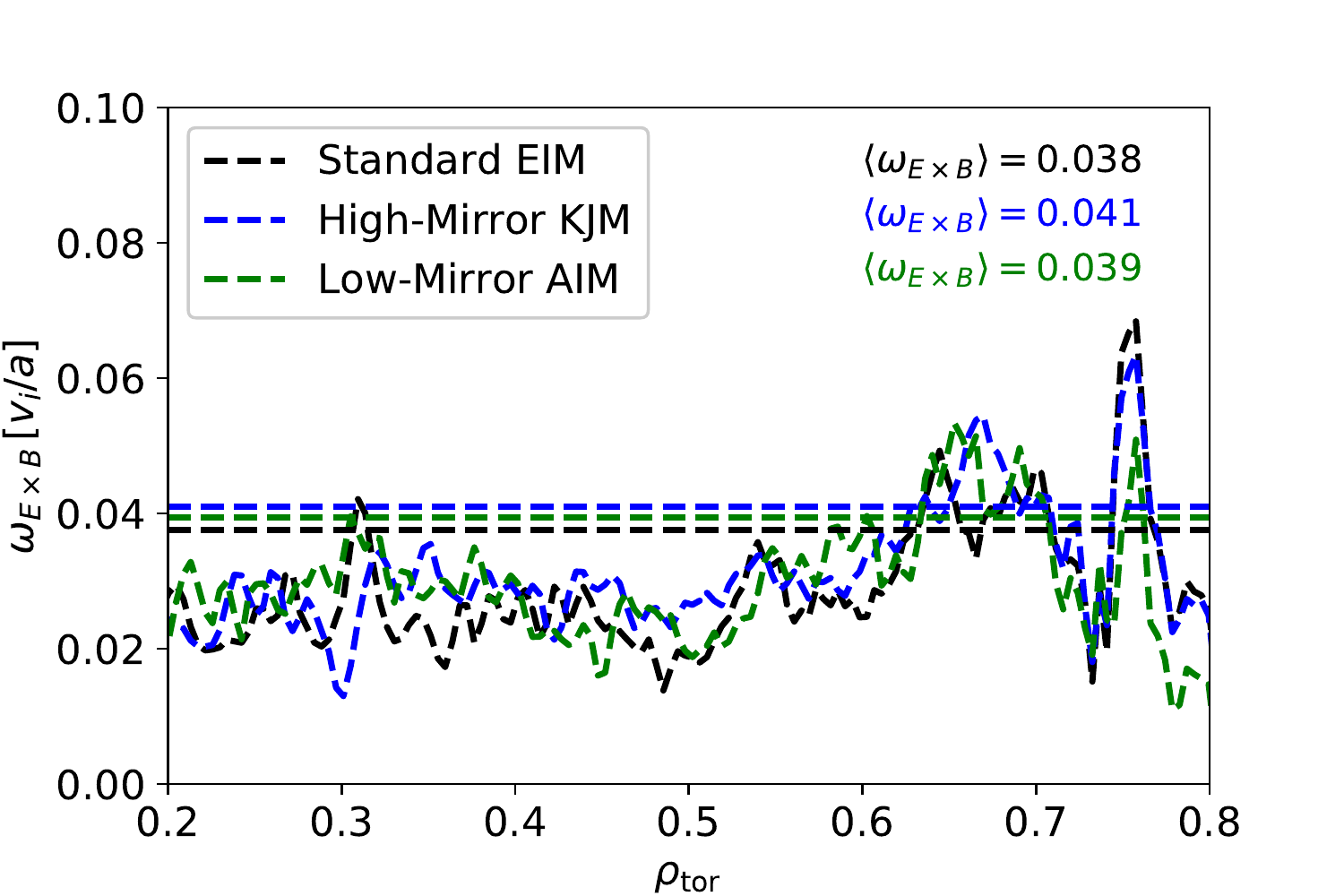} 
\end{tabular}
\caption{\label{fig:exb} Comparison of the radial profile of the $E \times B$ shearing rate. The horizontal dashed line shows its radial average whose value is also indicated.  }
\end{figure}

Finally, we compare the radial profile of $E \times B$ shear flows, defined as:
\begin{align}
\omega_{E \times B} = \frac{\rho_{\rm tor}}{B_{0}} \frac{\partial}{\partial \rho_{\rm tor}} \left( \frac{- \nabla \langle \phi_1 \rangle_{\rm{FS}}}{\rho_{\rm tor}}  \right),
\end{align}
since they are one of the important phenomena responsible for the saturation of turbulence~[\onlinecite{Diamond_2005}, \onlinecite{Helander_2011}]. We observe that the radial profiles of the shear flows (figure~\ref{fig:exb}) are practically identical (and smaller than the linear growth rate) for all the configurations, despite the higher heat flux in the Low-Mirror configuration.  Therefore, this result suggests that shear flows play a smaller role in the saturation of turbulence in the Low-Mirror configuration.

%%%%%%%
%%%%%%% 

To summarize the section, we showed that ITG turbulent transport depends on the magnetic configuration used in W7-X. This is contrary to what we observed in linear simulations.  In particular, we found that the Low-Mirror configuration produces more heat flux than both the Standard and the High-Mirror configurations. This happens despite similar levels of $E \times B$ shear flows, implying that the shear flows are less important for the saturation of turbulence in the Low-Mirror configuration. Nevertheless, we found that the spatial structure of the heat flux is very similar for all configurations, being localized in the regions of bad curvature,  although it is broader than in the linear case. The discrepancy between these results may be explained by the fact that the heat flux spectra are dominated by larger scale modes than in the corresponding linear simulations.

%%%%%%%%%%%%%%%%%%%%%%%%%%%%%%%%%%%%%%%%%%%%%%%%%%%%%%%%%%%%%%%
%%%%%%%%%%%%%%%%%%%%%%%%%%%%%%%%%%%%%%%%%%%%%%%%%%%%%%%%%%%%%%%
%%%%%%%%%%%%%%%%%%%%%%%%%%%%%%%%%%%%%%%%%%%%%%%%%%%%%%%%%%%%%%

\section{Comparison with radially local simulations~\label{discussion}}

%%%%%%%%%%%%%%%%%%%%%%%%%%%%%%%%%%%%%%%%%%%%%%%%%%%%%%%%%%%%%%%%
%%%%%%%%%%%%%%%%%%%%%%%%%%%%%%%%%%%%%%%%%%%%%%%%%%%%%%%%%%%%%%%%
%%%%%%%%%%%%%%%%%%%%%%%%%%%%%%%%%%%%%%%%%%%%%%%%%%%%%%%%%%%%%%%%

We have performed global simulations to study the effect of ITG turbulence on different magnetic geometries of W7-X and found that the Low-Mirror configuration produces more heat flux than both the Standard and the High-Mirror configurations. In this section, we want to investigate if a similar conclusion can be reached with a reduced - and less computationally demanding - model, such as a radially local one~[\onlinecite{Xanthopoulos_2014}, \onlinecite{Xanthopoulos_2016}].

\begin{figure}[b]
\begin{tabular} {c}
\includegraphics[width=0.95\linewidth,angle=0]{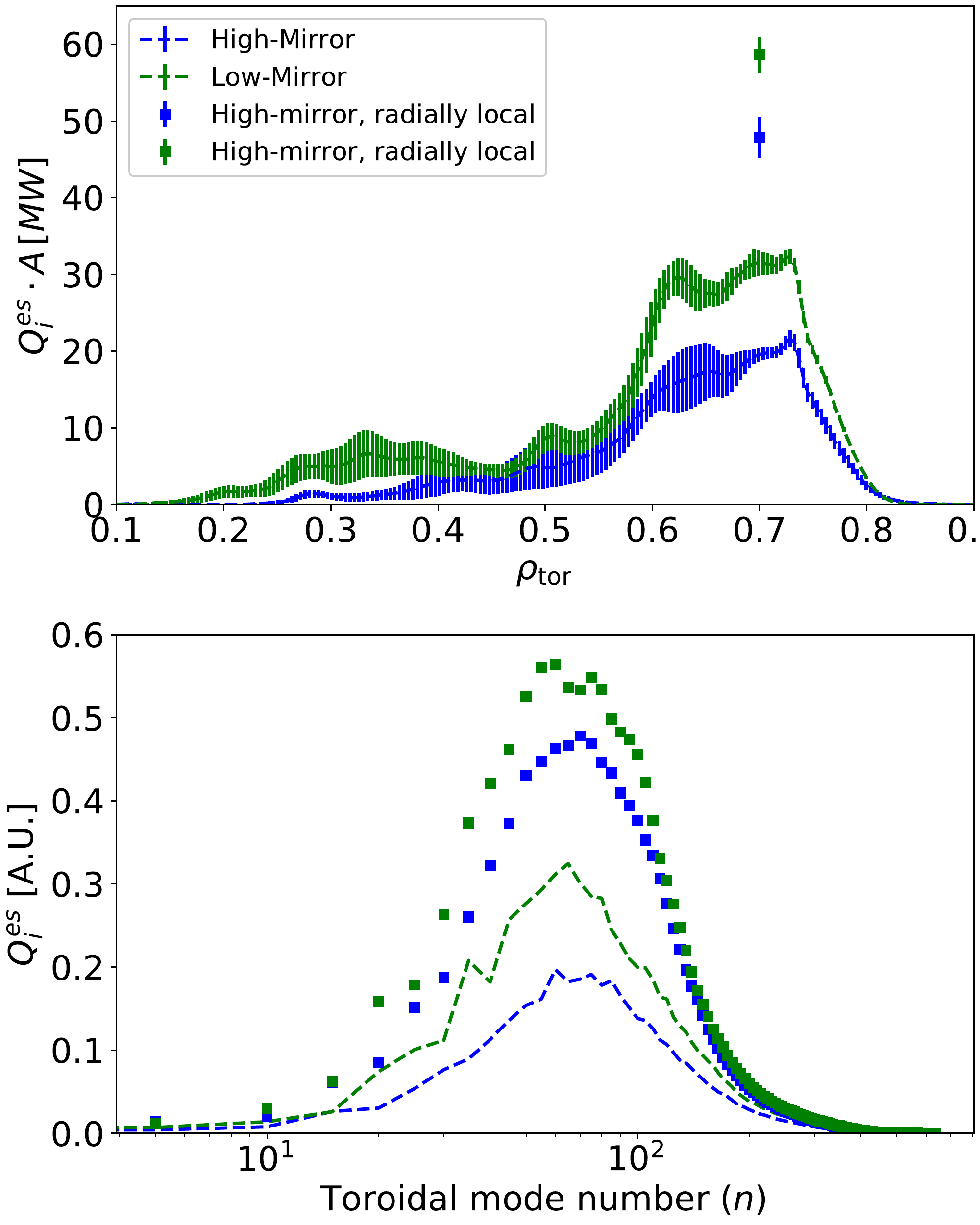} 
\end{tabular}
\caption{\label{fig:comp_local_global} Comparison between global and radially local simulations of  the heat fluxes (top) and their spectra at $\rho_{\rm tor} = 0.7$ (bottom) for the Low-Mirror and High-Mirror configurations.}
\end{figure}

For this purpose, we performed radially local simulations with GENE-3D. These simulations are conducted by selecting a flux-surface, employing constant geometry coefficients and profiles along the radial direction, but still keeping finite gradients and magnetic shear (as is usually done in local simulations in flux-tube codes), and using radially periodic boundary conditions. We selected the flux-surface where the heat flux was is at its maximum, i.e., $\rho_{\rm tor} = 0.7$, and performed two simulations: one for the High-Mirror and another for the Low-Mirror configuration. The resolution used is the same as in the global case, but the simulations are nevertheless cheaper (around $0.1$ million CPU-hours) since they need less time to reach a quasi-stationary state, and are also simpler to run because they do not require sources to maintain the equilibrium profiles. We note that in this position, the normalized system size is $\rho^{\star} = \rho_i / a \approx 1 /250 $, with $\rho_i$ being the ion Larmor radius.

We observe that the trend of the Low-Mirror configuration producing more heat flux than the High-Mirror is still present in the radially local simulations (figure~\ref{fig:comp_local_global}~(top)). In addition, we also observe a close agreement between global and radially local simulations regarding the shape of the heat flux spectra (figure~\ref{fig:comp_local_global}~(bottom)).  However, in both cases, the amplitudes of the fluxes are largely overestimated (up to a factor of $2.5$) in local simulations.

Therefore, these results indicate that for a qualitative study of the effect of the geometry on ITG turbulence, radially local simulations are sufficient. However, for a quantitative study, such as the validation of gyrokinetic simulations against experimental results, global simulations are generally expected to be necessary. Furthermore, global nonlinear simulations will still be required in cases where effects not captured in radially local simulations, such as the shear of the radial electric field~[\onlinecite{Mishchenko_2012}], are expected to play a key role in regulating turbulence.

%%%%%%%%%%%%%%%%%%%%%%%%%%%%%%%%%%%%%%%%%%%%%%%%%%%%%
%%%%%%%%%%%%%%%%%%%%%%%%%%%%%%%%%%%%%%%%%%%%%%%%%%%%%
%%%%%%%%%%%%%%%%%%%%%%%%%%%%%%%%%%%%%%%%%%%%%%%%%%%%%

\section{Conclusions\label{conclusions}}

%%%%%%%%%%%%%%%%%%%%%%%%%%%%%%%%%%%%%%%%%%%%%%%%%%%%%
%%%%%%%%%%%%%%%%%%%%%%%%%%%%%%%%%%%%%%%%%%%%%%%%%%%%%
%%%%%%%%%%%%%%%%%%%%%%%%%%%%%%%%%%%%%%%%%%%%%%%%%%%%%

In the present paper,  we performed linear and nonlinear global simulations of W7-X with GENE-3D
to understand how ITG turbulence depends on a particular configuration used in W7-X.

We found that the Low-Mirror configuration produces more turbulent transport than both the High-Mirror and the Standard configurations.  We observed that nonlinear simulations were necessary to correctly characterize how ITG turbulence is affected by the different magnetic geometries. The reason is that the most unstable mode in linear simulations is a large toroidal mode number, which does not contribute significantly to transport, and therefore it does not provide relevant information 
of the nonlinear system.  Finally, we found that a radially local model can capture the trends observed in the global model qualitatively. There were, however, quantitative differences in the amplitudes, showing that nonlinear global simulations are generally expected to be necessary to validate the gyrokinetic results against experimental observations in W7-X.

The results presented in this paper thus represent an ongoing effort in the gyrokinetic stellarator community in moving forward to perform nonlinear global simulations~[\onlinecite{Cole_2020,Sanchez_2020,Wang_2020}].
In future work, we would like to perform further studies, including a more comprehensive plasma description in the simulations. In particular, we would like to treat the electrons as a kinetic species, add a radial electric field and electromagnetic effects. The goal is to perform a gyrokinetic validation study for W7-X, which will then give us confidence in our model and could help the operation and design of new machines.

%%%%%%%%%%%%%%%%%%%%%%%%%%%%%%%%%%%%%%%%%%%%%%%%%%%%%%%%%%%%%%%%%%%
%%%%%%%%%%%%%%%%%%%%%%%%%%%%%%%%%%%%%%%%%%%%%%%%%%%%%%%%%%%%%%%%%%%
%%%%%%%%%%%%%%%%%%%%%%%%%%%%%%%%%%%%%%%%%%%%%%%%%%%%%%%%%%%%%%%%%%%

\begin{acknowledgements}

The authors would like to thank Y.~Turkin and J.~Geiger for providing, respectively,  the profiles and the VMEC equilibria used in this work. A.~Ba\~n\'on~Navarro would also like thank J.~P.~Martin~Collar for helpful suggestions and comments. Numerical simulations were performed at the MARCONI-Fusion supercomputer at CINECA, Italy, and at Cobra HPC system at the Max Planck Computing and Data Facility (MPCDF), Germany. 

\end{acknowledgements}

%%%%%%%%%%%%%%%%%%%%%%%%%%%%%%%%%%%%%%%%%%%%%%%%%%%%%%%%%%%%%%%%%%%
%%%%%%%%%%%%%%%%%%%%%%%%%%%%%%%%%%%%%%%%%%%%%%%%%%%%%%%%%%%%%%%%%%%
%%%%%%%%%%%%%%%%%%%%%%%%%%%%%%%%%%%%%%%%%%%%%%%%%%%%%%%%%%%%%%%%%%%

\section*{References}

\end{document}